\documentclass{ws-procs975x65}

\begin{document}

\title{Quantum Foam}

\author{Y. Jack Ng}
\address{Institute of Field Physics, Department of Physics and Astronomy,\\
University of North Carolina, Chapel Hill, NC 27599-3255, USA\\
E-mail: yjng@physics.unc.edu}

\maketitle

\abstracts{
Quantum foam, also known as spacetime foam, has its origin in quantum 
fluctuations of
spacetime.  Its physics is intimately linked to that of black holes and
computation.  Arguably it is the source of the holographic principle 
which severely limits how densely
information can be packed in space.
Various proposals to detect the foam are briefly discussed.
Its detection will provide us with a
glimpse of the ultimate structure of space and time.}

\section{Introduction}

Before last century, spacetime was regarded as nothing
more than a passive and static arena in which events took place.
Early last century, Einstein's general relativity
changed that viewpoint and promoted spacetime to an
active and dynamical entity.  
Nowadays many physicists also believe
that spacetime, like all matter and energy, undergoes quantum
fluctuations.  These quantum fluctuations make spacetime
foamy\footnote{For a brief review and a more complete list of references, 
see Ref. 1.} on small 
spacetime scales.

But how large are the fluctuations?
How foamy is spacetime?  Is there any theoretical evidence of 
quantum foam? 
And how can we detect quantum foam?
In what follows, we address these questions.  
The outline of this paper is as follows:
By analysing a gedanken experiment for spacetime measurement, we show,
in subsection 2.1, that spacetime fluctuations scale as the cube root of
distances or time durations.  In subsection 2.2, we show that this cube
root dependence is consistent with the holographic principle.  Subsection
2.3 is devoted to a comparison of this peculiar dependence with the 
well-known random-walk problem and other quantum gravity models.  Here we
also consider the cumulative effects of individual spacetime fluctuations.
In  
section 3, we discuss how quantum foam affects the physics of 
clocks and computation (subsection 3.1), and show that 
the physics of
spacetime foam is intimately connected to that of black holes
(subsection 3.2).  Just as there are uncertainties in spacetime
measurements, there are also uncertainties in energy-momentum measurements.
This topic of energy-momentum uncertainties
is given a brief treatment in section 4.  Some proposals to
detect quantum foam are considered in section 5.  One particular proposal
involving ultra-high energy cosmic ray events is discussed in
the Appendix.

Before we proceed, we should mention that the approach to the 
physics of 
quantum foam adopted here is very 
conservative: the only ingredients we use are quantum mechanics and general
relativity.  Hopefully, by 
considering only distances
(time durations) much larger than the
Planck length (time) or energies (momenta) much
smaller than Planck energy (momentum), a semi-classical 
treatment of gravity suffices and
a bona fide theory of quantum gravity is not needed.  We should also make
it clear at the outset that we make no assumptions on the high energy
regime of the ultimate quantum gravity theory.  We refrain from
speculating on violations of Lorentz invariance and the consequent 
systematically modified dispersion relations, involving a 
coefficient of {\it fixed} magnitude and {\it fixed} sign,
which many people believe are unavoidably induced by quantum gravity.
(In the terminology of Ref. 2, these quantum gravity effects are
called ``systematic'' effects.)  The only
quantum gravity effects we are concerned with in this paper are those
due to quantum fuzziness --- uncertainties involving {\it fluctuating}
magnitudes with {\it both} $\pm$ signs, perhaps like a fluctuation with a 
Gaussian distribution 
about zero.  (In the terminology of Ref. 2, these effects 
are called ``non-systematic'' effects.)

\section{Quantum Fluctuations of Spacetime}
\label{sec:quantum}

If spacetime indeed undergoes quantum
fluctuations, the fluctuations will show up when we measure a distance
(or
a time duration), in the form of uncertainties in the measurement.
Conversely, if in any distance (or time duration) measurement, we cannot
measure the distance (or time duration) precisely, we interpret this
intrinsic limitation to spacetime measurements as resulting from
fluctuations of spacetime.  

The question is: does spacetime undergo quantum fluctuations? And
if so, how large are the fluctuations?  To quantify the problem, let us
consider measuring a distance $l$. The question now is: how accurately
can
we measure this distance?  Let us denote by $\delta l$ the accuracy with
which we can measure $l$.
We will also refer to
$\delta l$ as the uncertainty or fluctuation of the distance $l$ for
reasons that will become obvious shortly.  We
will show that $\delta l$ has a lower bound and will use two ways
to calculate it.
Neither method is rigorous, but the fact that the
two very different methods yield the same result bodes well for the
robustness of the conclusion.

\subsection{Gedanken Experiment}
In the first method, we conduct a thought experiment to measure $l$.
The importance of carrying out spacetime measurements to find the
quantum fluctuations in the fabric of spacetime
cannot be
over-emphasized.  According to general relativity, coordinates do not
have
any intrinsic meaning independent of observations; a coordinate system
is defined only by explicitly carrying out spacetime distance
measurements.
Let us measure the distance between point A and point B.
Following Wigner\cite{wigner}, we put a clock
at A and a mirror at B.  Then the distance $l$ that we want to measure is
given
by the distance between the clock and the mirror.
By sending a light signal from the clock to the
mirror in a timing experiment, we can determine the distance $l$.  
However, the quantum uncertainty in the positions of
the clock and the mirror introduces an inaccuracy $\delta l$ in the
distance measurement.  We expect the clock and the mirror to contribute
comparable uncertainties to the measurement.
Let us concentrate on the clock and denote its mass
by $m$.  Wigner argued that if it has a linear
spread $\delta l$ when the light signal leaves the clock, then its position
spread grows to $\delta l + \hbar l (mc \delta l)^{-1}$
when the light signal returns to the clock, with the minimum at
$\delta l = (\hbar l/mc)^{1/2}$.  Hence one concludes that
\begin{equation}
\delta l^2 \gtrsim \frac{\hbar l}{mc}.
\label{sw}
\end{equation}
Thus quantum mechanics
alone would suggest using a massive clock to reduce the jittering of the
clock and thereby the uncertainty $\delta l$.  
On the other hand, according to
general relativity, a massive clock would distort the surrounding
space severely, affecting adversely the accuracy in the measurement of the
distance.  

It is here that we appreciate the importance of taking into account the 
effects of instruments in this thought-experiment.  Usually when one wants
to examine a certain a field (say, an electromagnetic field) one uses
instruments that are neutral (electromagnetically neutral) and massive
for, in that case, the effects of the instruments are negligible.  But
here in our thought-experiment, the relevant field is the gravitational
field.  One cannot have a gravitationally neutral yet massive set of 
instruments because the gravitational charge is equal to the mass
according to the principle of equivalence in general relativity.  Luckily
for us, we can now exploit this equality of the gravitational charge and
the inertial mass of the clock to eliminate the dependence on $m$ in the 
above inequality to promote Eq.~(\ref{sw}) to a (low-energy) quantum 
gravitational uncertainty relation.

To see this, let the clock be a
light-clock consisting of a spherical cavity of diameter $d$, 
surrounded by a mirror wall
of mass $m$, between which bounces a beam of light.  For
the uncertainty in distance measurement not to be greater than $\delta l$,
the clock must
tick off time fast enough that $d/c \lesssim \delta l /c$.  But $d$, the
size of the clock, must be larger than the Schwarzschild radius 
$r_S \equiv 2Gm/c^2$ of
the mirror, for otherwise one cannot read the time registered on the
clock.  From these two requirements, it follows that
\begin{equation}
\delta l \gtrsim \frac{Gm}{c^2}.
\label{ngvan}
\end{equation}
Thus general relativity alone
would suggest using a light clock to do the
measurement.
This result can also be derived in another way.  If the clock
has a radius $d/2$ (larger than its Schwarzschild radius $r_S$), 
then $\delta l$, the error in the distance measurement caused by the 
curvature generated by the mass of the clock, 
may be estimated by a calculation from the Schwarzschild 
solution.  The result is $r_S$ multiplied by a logarithm involving
$2r_S/d$ and $r_S/(l + d/2)$.  For $d >> r_S$, one finds 
$\delta l = \frac{1}{2}r_S \log{\frac{d + 2l}{d}}$ and hence 
Eq.~(\ref{ngvan}) as an order of magnitude estimate.

The product
of Eq.~(\ref{ngvan}) with Eq.~(\ref{sw}) yields
\begin{equation}
\delta l \gtrsim (l l_P^2)^{1/3} = l_P \left(\frac{l}{l_P}\right)^{1/3},
\label{nvd1}
\end{equation}
where $l_P = (\hbar G/c^3)^{1/2}$ is the Planck length.
(Note that the result is independent of the
mass of the clock and, thereby, one would hope,
of the properties of the specific
clock used in the measurement.)
The end result is as simple as it is strange
and appears to be universal: the uncertainty $\delta l$ in 
the measurement of the distance $l$ cannot be smaller than the cube root
of $l l_P^2$.\cite{ngvan1}
Obviously the accuracy of the
distance measurement is intrinsically limited by this
amount of uncertainty or
quantum fluctuation.  We conclude that
there is a limit to
the accuracy with which one can measure a distance;
in other words, we can never know the
distance $l$ to a better accuracy than the cube root of $l l_P^2$.
(Similarly one can show that we can never know a time duration
$\tau$ to a better accuracy than the cube root of $\tau t_P^2$, where
$t_P \equiv l_P/c$ is the Planck time.)\footnote{The spacetime
fluctuation translates into a metric fluctuation over a distance $l$
and a time interval $\tau$ given by $\delta g_{\mu \nu}$ greater than
$(l_P/l)^{2/3}, (t_P/\tau)^{2/3}$ respectively.  For a discussion of
the related light-cone fluctuations, see Ref. 5.}
Because the Planck length is so inconceivably short,
the uncertainty or intrinsic limitation to the accuracy in the
measurement
of any distance, though much larger than the Planck length,
is still very small.  For example,
in the measurement of a distance of
one kilometer, the uncertainty in the distance is to an atom as
an atom is to a human being.

\subsection{The Holographic Principle}
Alternatively we can estimate $\delta l$ by applying the holographic
principle.\cite{found,stfoam}
In essence, the holographic principle\cite{wbhts}
says that although the world
around us appears to have three spatial dimensions, its contents can
actually
be encoded on a two-dimensional surface, like a hologram.  To be more
precise,
let us consider a spatial region measuring $l$ by $l$ by $l$.  According
to the
holographic principle, the number of degrees of freedom that this cubic
region
can contain is bounded by the surface area of the region in Planck units,
i.e.,
$l^2 / l_P^2$, instead of by the volume of the region as one may naively
expect.  This principle is strange and counterintuitive, but 
is supported by black hole physics in conjunction with
the laws of thermodynamics, and it is embraced by both string theory and
loop gravity, two top contenders of quantum gravity theory.
So strange as it may be, let us now apply the holographic
principle to
deduce the accuracy with which one can measure a distance.

\begin{figure}[ht]
\centerline{\epsfxsize=2.6in\epsfbox{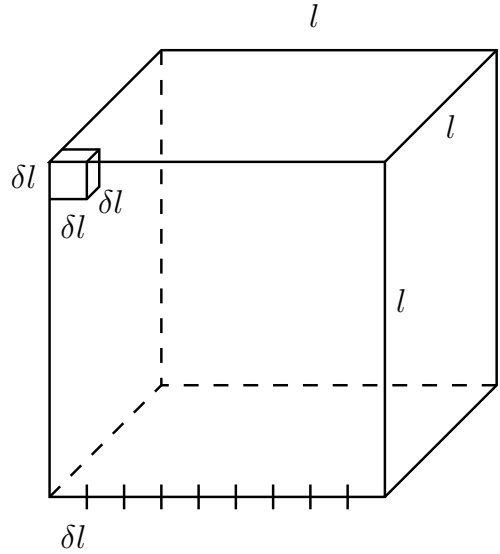}}
\caption{
Partitioning a big cube into small cubes.  The big cube represents a
region of space measuring $l$ by $l$ by $l$.  The small cubes represent
the smallest physically-allowed cubes measuring $\delta l$ by $\delta l$
by
$\delta l$ that can be lined up to measure the length of each side of the
big
cube.  Strangely, the size of a small cube is {\it not} universal, but   
depends on the size of the big cube.  A simple argument
based on this construction leads to the holographic principle.
\label{fig1}
}
\end{figure}

First,
imagine partitioning the big cube into small cubes [see Fig.~\ref{fig1}].
The small cubes so constructed should
be as small as physical laws allow so that we can associate one degree of
freedom with each small cube.   In other words, the number of
degrees of freedom that the region can hold is given by the number of
small
cubes that can be put inside that region.
But how small can such cubes be?
A moment's thought tells us that each side of a small cube
cannot be smaller than the accuracy
$\delta l$ with which we can measure each side $l$ of the big cube.
This can be easily shown by applying the method of contradiction:  assume
that we can construct small cubes each of which has sides less than
$\delta l$.
Then by lining up a row of such small cubes along a side of
the big cube from end to end, and by counting the number of such small
cubes,
we would be able
to measure that side (of length $l$) of the big cube
to a better accuracy than $\delta l$.  But, by
definition, $\delta l$ is the best accuracy with which we can measure
$l$.  The
ensuing contradiction is evaded by the realization that each of the
smallest
cubes (that can be put inside the big cube) measures $\delta l$ by
$\delta l$ by
$\delta l$.  Thus, the number of degrees of freedom in the region
(measuring $l$ by $l$ by $l$) is given by $l^3 / \delta l^3$,
which,
according to the holographic principle, is no more than $l^2 / l_p^2$.
It follows
that $\delta l$  is bounded (from below) by the cube root of $l l_P^2$,
the same result as found
above in the gedanken experiment argument.  Thus, to the extent that the
holographic principle is correct, spacetime indeed fluctuates, forming
foams of 
size $\delta l$ on the scale of $l$.  Actually,
considering the fundamental nature of spacetime and the ubiquity of
quantum fluctuations, we should reverse the
argument and then we will come to the conclusion that the
``strange'' holographic principle has its
origin in quantum fluctuations of spacetime.\footnote{Recently, Scardigli and
Casadio\cite{casadio} claim that the expected holographic scaling seems 
to hold only in
(3+1) dimensions and only for the ``generalized uncertainty principle'' 
found above for $\delta l$.}

\subsection{Quantum Gravity Models}
The consistency of the uncertainties in distance measurements
with the holographic principle is reassuring.  But the dependence of the
fluctuations
in distance on the cube root of the distance is still perplexing.  To
gain further insight into this strange state of affairs,
let us compare this peculiar dependence on distance with the
well-known one-dimensional random-walk problem.  For a random walk of
steps of equal size, with each step equally likely to either direction, 
the root-mean-square deviation from the mean is given by the size of each
step multiplied by the square
root of the number of steps.
It is now simple to concoct a random-walk
model\cite{AC,dio89} for the
fluctuations of distances in quantum gravity.  Consider a distance $l$,
which we partition into $l/l_P$ units each of length $l_P$.
In the random-walk model of quantum gravity, $l_P$ plays the role of the
size of each
step and $l/l_P$ plays the role of the number of steps.  
The fluctuation in distance $l$ is given by $l_P$ times the square root of
$l/l_P$, which comes out to the square root of $l l_P$.  
This is much bigger than the cube root of
$l l_P^2$, the fluctuation in distance measurements found above.

\begin{figure}[ht]
\centerline{\epsfxsize=3.8in\epsfbox{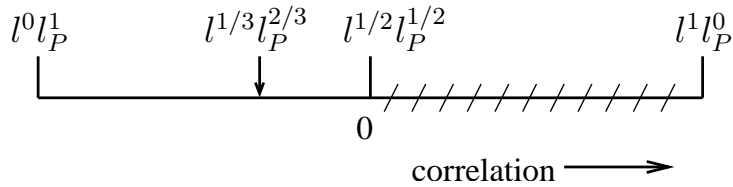}}
\caption{
Lower bounds on $\delta l$ for the various quantum gravity models.
The fluctuation of the distance $l$ is given by the
sum of $l/l_P$ fluctuations each by plus or minus $l_P$.
Spacetime foam appears to choose a small anti-correlation
(i.e., negative correlation) between
successive fluctuations, giving
a cube root dependence in the number
$l/l_p$ of fluctuations for the total fluctuation 
of $l$ (indicated by the arrow).  It falls between
the two extreme cases of complete randomness, i.e., zero
(anti-)correlation (corresponding to $\delta l \sim l^{1/2}l_P^{1/2}$) and
complete anti-correlation (corresponding to $\delta l \sim l_P$).  Quantum
gravity models corresponding to positive correlations between successive   
fluctuations (indicated by the hatched portion) are observationally 
ruled out.
\label{fig2}
}
\end{figure}

The following interpretation of the dependence of $\delta l$ on the cube
root of $l$
now presents itself.  As in the random-walk model,
the amount of fluctuations in the distance
$l$ can be thought of as an accumulation of the
$l/l_P$ individual fluctuations each by an amount plus or minus $l_P$.
But, for this case, the individual fluctuations cannot be completely 
random (as opposed to the random-walk model); actually
successive fluctuations must be somewhat anti-correlated
(i.e., a plus fluctuation is slightly more likely followed by a minus
fluctuation and vice versa),
in order that together they produce a total fluctuation less
than that in the random-walk model.  This small amount of
anti-correlation between successive fluctuations (corresponding to
what statisticians call fractional Brownian motion with
self-similarity parameter $\frac{1}{3}$)
must be due to quantum gravity effects.  
Since the cube root dependence
on distance has been shown to be consistent with the 
holographic principle, 
we will, for the rest of this subsection, refer to this case that 
we have found 
(marked by an arrow in Fig.~\ref{fig2}) as the 
holography model.

On the other hand, if
successive fluctuations are completely anti-correlated, i.e., a
fluctuation by plus $l_P$
is followed by a fluctuation by minus $l_P$ which is succeeded by plus
$l_P$ etc. in
the pattern $+-+-+-+-+-...$, then the fluctuation of a distance $l$ is
given by the minuscule $l_P$,\cite{mis73}
independent of the size of the distance.  Thus the holography model
falls between the two extreme cases of complete randomness
(square root of $l l_P$) and complete anti-correlation ($l_P$).  
For completeness, we mention that {\it a priori} there are also models with 
correlating successive fluctuations.  But these models yield unacceptably
large fluctuations in distance and time duration measurements ---
we will see below that these models (corresponding to the hatched line to 
the right of the random-walk model shown in Fig.~\ref{fig2})
have already been observationally ruled out.  

Let us now examine the cumulative effects\cite{NCvD}
of spacetime fluctuations over
a large distance.
Consider a distance $l$, 
and divide it into $l/ \lambda$ equal
parts each of which has length $\lambda$. 
If we start with $\delta \lambda$ from each part, the question is how do
the $l/ \lambda$ parts
add up to $\delta l$ for the whole distance $l$.  In other words, we want
to find
the cumulative factor $\mathcal{C}$ defined by
\begin{equation}
\delta l = \mathcal{C}\, \delta \lambda,
\label{cf1/2.1}
\end{equation}
For the holography model, since $\delta l \sim l^{1/3} l_P^{2/3} = l_P 
(l/l_P)^{1/3}$ and
$\delta \lambda \sim {\lambda}^{1/3} l_P^{2/3} =
l_P (\lambda/l_P)^{1/3}$, the result is
\begin{equation}
\mathcal{C} = \left(\frac{l}{\lambda}\right)^{1/3}.
\label{cf}
\end{equation}

For the random-walk model, the cumulative factor is given by 
$\mathcal{C} = (l/ \lambda)^{1/2}$; for the model corresponding
to complete anti-correlation, the cumulative factor is
$\mathcal{C} = 1$, {\it independent} of $l$.
Let us note that, for all
quantum gravity models (except for the physically disallowed
model corresponding to complete correlation between successive fluctuations),
the cumulative factor is {\it not} linear in $(l/\lambda)$, i.e.,
$\frac{\delta l}{\delta \lambda} \neq \frac{l}{\lambda}$.  (In fact, it is
much smaller than $l/\lambda$).
The reason for this is obvious: the $\delta \lambda$'s from the $l/ \lambda$
parts in $l$ do {\it not} add coherently.  It makes no sense, e.g., to say,
for the completely anti-correlating model, that $\delta l \sim \delta \lambda
\times l/\lambda \gtrsim l_P l/\lambda$ because it is inconsistent to use the
completely anti-correlating model for $\delta \lambda$ while using the
completely correlating model for the cumulative factor.

Note that the above discussion on cumulative effects
is valid for {\it any} $\lambda$ between
$l$ and $l_P$, i.e., it does not matter how one partitions the distance
$l$.  In particular, for our holography model, one can choose to partition 
$l$ into units of Planck length $l_P$, the 
smallest physically meaningful length.  Then (for $\lambda = l_P$)
using $\delta l_P \sim
l_P^{1/3} \times l_P^{2/3} = l_P$, one recovers 
$\delta l \sim (l/l_P)^{1/3} \times l_P = l^{1/3} l_P^{2/3}$, with the 
dependence on the cube root of $l$ being due to a small amount of 
anti-correlation between successive fluctuations as noted above.
The fact that we can choose $\lambda$ as small as the Planck length in the 
partition indicates that, in spite of our earlier disclaimer, it may even be 
meaningful to consider, in the semi-classical framework we are pursuing,
fluctuations of distances close to the Planck length.  

Now that we know where the holography model stands among the quantum 
gravity models, we will restrict ourselves
to discuss this model only for the rest of the paper.

\section{From Spacetime Foam to Computers to Black Holes}
So far there is no experimental evidence for spacetime foam, and, as we
will show shortly, no direct evidence is expected in the very near
future.  In view of this lack of
experimental evidence, we should at least look for theoretical
corroborations (aside
from the ``derivation" of the holographic principle discussed above).
Fortunately such corroborations do exist --- in the sector of black hole
physics (this should not come as a surprise to the experts).  To
show that, we have to make a small detour to consider clocks and
computers\cite{ng,barrow} first.

\subsection{Clocks and Computers}
Consider a clock (technically, a simple and ``elementary'' clock, not
composed of smaller
clocks that can be used to read time separately or sequentially), capable
of resolving time to an accuracy of $t$, for a period of
$T$ (the running time or lifetime of the clock).
Then bounds on the resolution time and the lifetime of the clock can be
derived by following an argument very similar
to that used above in the analysis of the gedanken experiment to measure
distances.
Actually, the two arguments are so similar that one can identify the
corresponding quantities.  [See Table.]

\begin{table}[bt]
\tbl{The corresponding quantities in the discussion of distance
measurements (first column),
time duration measurements (second column), clocks
(third column), and computers (fourth column) appear
in the same row in the following Table.}
{\footnotesize
\begin{tabular}{llll} \hline
distance & time duration & clocks &
computers \\
measurements & measurements & & \\   \hline\hline
distance uncertainty & time duration
& resolution & reciprocal of \\
divided by speed & uncertainty & time & computation \\
of light ($\delta l/c$) & ($\delta \tau$) & ($t$) & speed ($1/\nu$)\\  
\hline
distance & time duration & running &
number of bits \\
divided by speed & ($\tau$) & time & divided by compu- \\
of light ($l/c$) & & ($T$) & tation speed ($I/\nu$)\\  \hline
\end{tabular} }
\vspace*{13pt}
\end{table}

For the discussion of clocks,
one argues that at the end of the running time $T$, the linear spread
of the clock (of mass $m$)
grows to $\delta l \gtrsim (\hbar T/m)^{1/2}$.  But the position
uncertainty due to the act of time measurement must be smaller than the
minimum wavelength of the quanta used to read the clock: 
$\delta l \lesssim ct$,
for the entire period $T$.  It follows that\cite{ng}
\begin{equation}
t^2 \gtrsim \frac{\hbar T}{mc^2}, 
\label{clock1}
\end{equation}
which is the analogue of Eq.~(\ref{sw}).  On the other hand, for the clock
to be able to resolve time interval as small as $t$, the cavity of the 
light-clock must be small enough such that $d \lesssim ct$; 
but the clock must also be larger
than the Schwarzschild radius $2Gm/c^2$ so that the time registered by 
the clock can be read off at all.  These two requirements are satisfied with
\begin{equation}
t \gtrsim \frac{Gm}{c^3},
\label{clock3}
\end{equation}
the analogue of Eq.~(\ref{ngvan}).
One can combine the above two equations to give\cite{ng}
\begin{equation}
T/ t^3 \lesssim t_P^{-2} = \frac{c^5}{\hbar G},
\label{clock2}
\end{equation}
which relates clock precision to its lifetime.  
Numerically, 
for example, for a femtosecond ($10^{-15}$ sec) precision, the bound on the
lifetime of a simple clock
is $10^{34}$ years.

One can easily translate
the above relations for clocks into useful relations for a simple
computer
(technically, it refers to a computer designed to perform highly serial
computations, i.e., one that is not divided into subsystems computing in
parallel).  
Since the resolution
time $t$ for clocks is the smallest time interval relevant in the
problem, the fastest
possible processing frequency is given by its reciprocal, i.e., $1/t$.
Thus if $\nu$
denotes the clock rate of the computer, i.e., the number of operations
per bit per unit
time, then it is natural to identify $\nu$ with $1/t$.  
To identify the number $I$
of bits of information in the memory space of a simple computer, we recall
that the running time $T$ is the longest time interval relevant in the
problem.  Thus,
the maximum number of steps of information processing is given by the
running time divided by the resolution time, i.e., $T/t$.
It follows that one can identify the number $I$ of bits of the
computer with $T/t$.\footnote{One can think of a tape of length $cT$ 
as the memory space, partitioned into bits each of length $ct$.}
In other words, 
the translations from the case of clocks to the case of computers
consist of substituting the clock rate of computation 
for the reciprocal of
the resolution time, and substituting the number of bits for the running
time divided by the resolution time.
[See Table.]  The bound on the precision and lifetime of a clock 
given by Eq.~(\ref{clock2}) is
now translated into a bound on the rate of computation and number of bits
in the computer, yielding 
\begin{equation}
I \nu^2 \lesssim \frac{c^5}{\hbar G} \sim 10^{86} /sec^2.
\label{computer}
\end{equation}
The latter bound is intriguing: it requires the product of the
number of bits and the square of the computation rate for
{\it any} simple computer to be less than the square of the
reciprocal of Planck time,\cite{ng}
which depends on relativistic quantum gravity
(involving $c$, $\hbar$, and $G$).  This relation links
together our concepts of information/computation, relativity, gravity,
and quantum uncertainty.  
Numerically, the
computation bound is about seventy-six orders of magnitude above
what is available for a current lap-top computer performing ten billion
operations per second on ten billion bits, for which 
$I \nu^2 \sim 10^{10}/s^2$.

\subsection{Black Holes}

Now we can apply what we have learned about clocks and computers
to black holes.\cite{ng,barrow}
Let us consider using a black hole to measure time.
It is reasonable to use
the light travel time around the black hole's horizon as the resolution
time of the clock,
i.e., $t \sim \frac{Gm}{c^3} \equiv t_{BH}$, then 
from Eq.~(\ref{clock1}), one immediately finds that
\begin{equation}
T \sim \frac{G^2 m^3}{\hbar c^4} \equiv T_{BH}.
\label{Hawking}
\end{equation}
We have just recovered
Hawking's result for black hole lifetime!  

Finally, let us consider using a black hole to do computations.  This may
sound like a ridiculous proposition.  But if we believe that black holes
evolve according to quantum mechanical laws, it is possible, at least in
principle, to program black holes to perform computations that
can be read out of the fluctuations in the Hawking black hole radiation.
How large is the memory space of a black hole computer, and 
how fast can it compute?
Applying the results for computation derived above, we readily find
the number of bits in the memory space of a black hole computer, given by
the lifetime of the black hole divided by its resolution time as a clock,
to be 
\begin{equation}
I = \frac{T_{BH}}{t_{BH}} \sim \frac{m^2}{m_P^2} \sim \frac{r_S^2}{l_P^2},
\label{bhcomputer1}
\end{equation}
where $m_P = \hbar/(t_P c^2)$ is the Planck mass, $m$ and $r_S^2$ denote the 
mass and event
horizon area of the black hole respectively.
This gives the number of bits $I$ as the event horizon area in Planck units,
in agreement with
the identification of a black hole entropy.  Furthermore,
the number of operations per unit time for a
black hole computer is given by 
\begin{equation}
I \nu \sim mc^2/\hbar,
\label{bhcomputer2}
\end{equation}
its energy divided by Planck's constant,
in agreement with the result
found by Margolus and Levitin, and
by Lloyd\cite{Lloyd} (for the ultimate limits to
computation).  It is curious that all the bounds on computation
discussed above are saturated by black hole computers.  Thus one can even
say that once they are programmed to do computations, black holes are the
ultimate simple computers.

\begin{figure}[ht]
\centerline{\epsfxsize=3.8in\epsfbox{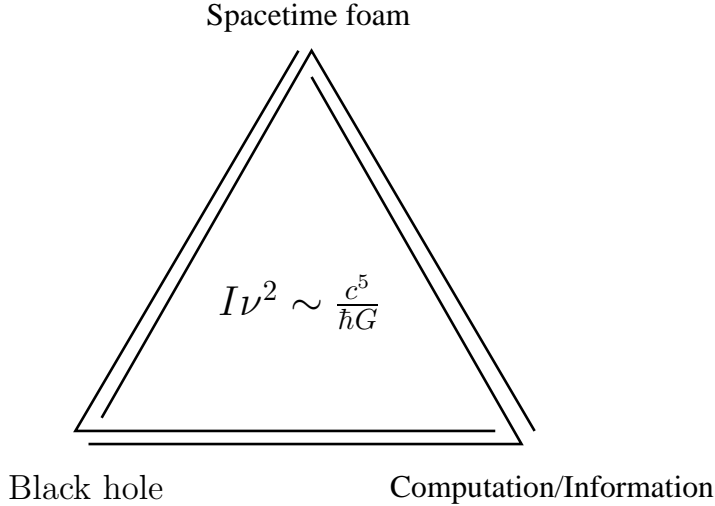}}
\caption{
The quantum foam-black hole-computation/information triangle.  
At the center of the triangle is the {\it universal}
relation: $I \nu^2 \sim  c^5/\hbar G$,
where $I$ is the number of bits
in the memory space, and $\nu$ is the clock rate of computation of a  
black hole computer.
This relation is a combined
product of the physics behind spacetime foam,
black holes, and computation/information.
\label{fig3}
}
\end{figure}

All these results reinforce the conceptual interconnections of the
physics underlying spacetime foam, black holes, and 
computation.
It is intersting that these three subjects
share such intimate bonds and are brought together here [see Fig.~\ref{fig3}].
The internal consistency of the physics we have uncovered
also vindicates the simple (some would say overly simple) 
arguments we present in section~\ref{sec:quantum}
in the derivation of the limits to spacetime measurements.  

\section{Energy-Momentum Uncertainties}

Just as there are uncertainties in spacetime measurements, there are
also uncertainties in energy-momentum measurements due to
spacetime foam effects.  Thus there is a limit to how accurately we
can measure and know the energy and momentum of 
a system.\cite{ngvan1}
Imagine sending
a particle of momentum $p$ to probe a certain structure of spatial extent 
$l$ so that $p \sim \hbar/l$.
It follows that $\delta p \sim (\hbar/l^2) \delta l$.  Spacetime fluctuations
$\delta l \gtrsim l (l_P/l)^{2/3}$ can
now be used to give
\begin{equation}
\delta p = \beta p \left(\frac{p}{m_P c}\right)^{2/3},
\label{dp}
\end{equation}
where {\it a priori} $\beta \sim 1$.
The corresponding statement for energy uncertainties is
\begin{equation}
\delta E = \gamma E \left(\frac{E}{E_P}\right)^{2/3},
\label{de}
\end{equation}
where $E_P = m_P c^2$ is the Planck energy and {\it a priori} 
$\gamma \sim 1$.
We emphasize that all the uncertainties take on $\pm$ sign with equal
probability (most likely, a Gaussian distribution about zero).
Thus at energy-momentum far below the Planck scale,
the energy-momentum
uncertainties are very small, suppressed by a fractional
(two-thirds) power of the Planck energy-momentum.  
(For example, the uncertainty in the energy of a particle of
ten trillion electron-volts
is about a thousand electron-volts.)  

Energy-momentum uncertainties affect both the energy-momentum conservation
laws and the dispersion relations.  Energy-momentum is
conserved up to energy-momentum uncertainties due to quantum foam effects,
i.e., $\Sigma (p_i^{\mu} + \delta p_i^{\mu}$)
is conserved, with $p_i^{\mu}$ being the average values 
of the various energy-momenta.  On the other hand
the dispersion relation is now generalized to read
\begin{equation}
E^2 - p^2c^2 - \epsilon p^2c^2 \left({pc \over E_P}\right)^{2/3} = m^2c^4,
\label{moddisp}
\end{equation}
for high energies with $E \gg mc^2$.  {\it A priori} we expect 
$\epsilon \sim 1$ and is independent of $\beta$ and $\gamma$.  But due to our 
present ignorance of quantum gravity, we are not in a position to make any
definite statements.  In fact, it is possible that 
$\epsilon = 2( \beta - \gamma)$, which would be the case if the modified
dispersion relation is given by $(E + \delta E)^2 -(p + \delta p)^2c^2 =
m^2c^4$.

The modified dispersion relation discussed above has an interesting
consequence for the speed of light.\cite{Ellis,NLOvD}
Applying Eq.~(\ref{moddisp})
to the massless photon yields
\begin{equation}
E^2 \simeq c^2p^2 + \epsilon E^2 \left(\frac{E}{E_P}\right)^{2/3}.
\label{gamd}
\end{equation}
The speed of (massless) photon
\begin{equation}
v = \frac{\partial E}{\partial p} \simeq c
\left( 1 +\frac{5}{6} \epsilon 
\frac{E^{2/3}}{E_P^{2/3}}\right),
\label{gams}
\end{equation}
becomes energy-dependent and fluctuates around c.
For example, 
a photon of ten trillion electron-volt energy has a speed 
fluctuating about $c$ by several centimeters per second.

\section {Spacetime Foam Phenomenology}

Because the Planck length $l_P \sim 10^{-33}$~cm is so minuscule, the Planck 
time $t_P \sim 10^{-44}$~sec so
short, and the Planck energy $E_P \sim 10^{28}$~eV so high,
spacetime foam effects, suppressed by Planck scales, are
exceedingly small.
Accordingly, they are very hard to detect.  
The trick will be to find ways to
amplify the small effects.\cite{br}

\subsection{Phase Incoherence of Light from Extra-galactic Sources}

One way to amplify the minute effects is to add up many such
effects, like collecting many small raindrops to fill a
reservoir.  Consider light coming to us from extragalactic
sources.  Over one wavelength,
the phase of the light-waves advances by $2 \pi$; but due to
spacetime foam effects, this phase fluctuates by a small amount.
The idea is that the fluctuation of the phase over
one wavelength is extremely small, but light from distant galaxies
has to travel a distance of many wavelengths.  It is
possible that over so many wavelengths, the fluctuations can
cumulatively add up to a detectable level
at which point the phase coherence for the light-waves is lost.
Loss of phase coherence would mean the loss of interference
patterns.  Thus
the strategy is to look for the blurring of images of distant
galaxies in powerful telescopes like the Hubble Space Telescope.
This technique to detect spacetime foam was proposed by
Lieu and Hillman\cite{lie03}, and elaborated by Ragazzoni and his 
collaborators\cite{rag03}.

The proposal deals with the phase
behavior of radiation with wavelength $\lambda$ received from a celestial
source located at a distance $l$ away.  Fundamentally, the wavelength defines
the minimum length scale over which physical quantities such as phase and
group velocities (and hence dispersion relations) can be defined.  Thus, the
uncertainty in $\lambda$ introduced by spacetime foam is the starting point
for this analysis.
A wave will travel a distance
equal to its own wavelength $\lambda$ in a time $t = \lambda / v_g$
where $v_g$ is the group velocity of propagation, and the
phase of the wave consequently changes by an amount
\begin{equation}
\phi = 2 \pi {v_p t\over \lambda} = 2 \pi {v_p\over v_g},
\label{phase1}
\end{equation}
(i.e., if $v_p = v_g, \phi = 2 \pi$)
where $v_p$ is the phase velocity of the light wave.  
Quantum gravity fluctuations, however, introduce random uncertainties
into this phase which is simply
\begin{equation}
\delta \phi =  2 \pi \, \delta\!\!\left({v_p\over v_g}\right).
\label{phase2}
\end{equation}

Due to quantum fluctuations of
energy-momentum\cite{ngvan1} and the modified dispersion relations,
we obtain
\begin{equation}
\delta\! \left( {v_p \over v_g}\right) \sim \pm \left({E\over E_P}
\right)^{2/3} = \pm \left({l_P \over \lambda}\right)^{2/3},
\label{deltav}
\end{equation}
where we have used $v_p = E/p$ and $v_g = dE / dp$, 
and 
$E/E_P = l_P / \lambda$.  
We emphasize that
this may be either an incremental advance or a retardation in the phase.

In travelling over the macroscopically large distance, $l$, from source
to observer an electromagnetic wave is continually subjected to random,
incoherent spacetime fluctuations.
Therefore, by our previous argument given in subsection 2.3, the 
cumulative statistical phase dispersion is 
$\Delta \phi = \mathcal{C} \delta \phi$ with the cumulative factor
$\mathcal{C} = (l/\lambda)^{1/3}$, that is
\begin{equation}
\Delta \phi =   2 \pi a \left({l_P\over
\lambda}\right)^{2/3} \left({l\over \lambda}\right)^{1/3}
= 2 \pi a {l_P^{2/3} l^{1/3} \over \lambda},
\label{delphi}
\end{equation}
where $a \sim 1$.  (This is our fundamental disagreement\cite{NCvD}
with Lieu and
Hillman who assume that the microscale fluctuations induced by quantum foam
into the phase of electromagnetic waves are coherently magnified by the 
factor $l/ \lambda$ rather than $(l/\lambda)^{1/3}$.)
Thus even the
active galaxy PKS1413+135, an example used by Lieu and Hillman,
which is more than four billion light years from Earth, is not
far enough to make the light wave front noticeably distorted.
A simple calculation\cite{NCvD} 
shows that, over four billion light years, the
phase of the light waves fluctuates only by one billionth
of what is required to lose the sharp ring-like interference
pattern around the galaxy which,
not surprisingly, is observed\cite{per02} by the 
Hubble Telescope.  This example illustrates the degree of
difficulty which one has to overcome to detect spacetime foam.
The origin of the difficulty can be traced to the incoherent
nature of the spacetime fluctuations (i.e., the anticorrelations 
between successive fluctuations).

But not all is lost with
Lieu and Hillman's proposal.  One can check that the proposal can be used
to rule out\cite{NCvD}, if only marginally, the random-walk model of
quantum gravity, which would (incorrectly) predict a large enough phase
fluctuation for light from PKS1413+135 to lose phase
coherence, contradicting evidence of diffraction patterns from
the Hubble Telescope observation.  It follows that 
models corresponding
to correlating successive fluctuations are also ruled out.

\subsection{High Energy $\gamma$ Rays from Distant GRB}

For another idea to detect spacetime foam, let us recall that,
due to quantum fluctuations of spacetime, the
speed of light fluctuates around $c$ and the fluctuations increase
with energy.  Thus
for photons (quanta of light) emitted simultaneously from a distant
source coming towards our detector, we expect an energy-dependent
spread in their arrival times.  To maximize the spread in arrival
times, we should look for energetic photons from distant sources.
High energy gamma rays from distant gamma ray bursts\cite{Ellis} fit the
bill.  So the idea is to look for a noticeable spread in arrival
times for such high energy gamma rays from distant gamma ray
bursts.  This proposal was first made by G. Amelino-Camelia 
et al.\cite{Ellis} in another context.

To underscore the importance of using the correct cumulative factor
to estimate the spacetime foam effect, let us first proceed in a naive 
manner. 
At first sight, the fluctuating speed of light would {\it seem} to 
yield\cite{NLOvD} an energy-dependent spread
in the arrival times of photons of the {\it same} energy $E$ given by
$\delta t \sim |\epsilon|t (E/E_P)^{2/3}$,
where $t$ is the average overall time of travel from the photon source.
Furthermore, the modified energy-momentum dispersion relation would seem to
predict
time-of-flight differences between simultaneously-emitted photons of
different energies, $E_1$ and $E_2$, given by
$\delta t \simeq \epsilon t (E_1^{2/3} - E_2^{2/3})/E_P^{2/3}$.
But these results for the spread of arrival times of photons are
{\it not} correct, because we have inadvertently 
used $l/\lambda \sim Et/\hbar$ as the cumulative factor instead of the 
correct factor $(l/\lambda)^{1/3} \sim (Et/\hbar)^{1/3}$.  Using the
correct cumulative factor, we get a much smaller $\delta t \sim 
t^{1/3} t_P^{2/3}$ for the spread in arrival time of the
photons of the same energy.
Thus the result is that the time-of-flight differences
increase only with the cube root of the
average overall time of travel from the gamma ray bursts to our
detector, leading to a time spread too small to be detectable.\cite{br}

\subsection{Interferometry Techniques}
Suppressed by the extraordinarily short Planck length, fluctuations
in distances, even large distances, are very small.  So,
to measure such fluctuations, what one needs is an instrument
capable of accurately measuring fluctuations in length over
long distances.  Modern gravitational-wave interferometers,
having attained extraordinary sensitivity, come to mind.  The idea
of using gravitational-wave interferometers to measure the
foaminess of spacetime was proposed by 
Amelino-Camelia\cite{AC} and elaborated by the
author and van Dam\cite{found}.  Modern gravitational-wave interferometers 
are sensitive to changes in distances to an
accuracy better than $10^{-18}$ 
meter.  To attain such sensitivity, interferometer researchers
have to contend with many different noises, the enemies of
gravitational-wave research, such as thermal noise, seismic noise,
and photon shot noise.  To this list of noises that infest an
interferometer, we now have to add the faint yet ubiquitous
noise from spacetime foam.  In other words, even after one has
subtracted all the well-known noises, there is still the noise from
spacetime fluctuations left in the read-out of the
interferometer.

The secret of this proposal to detect spacetime foam lies in the
existence of another length scale\cite{AC} available in this particular
technique, in addition to the minuscule Planck length.
It is the scale provided by the
frequency $f$ of the interferometer bandwidth.  What is important is
whether the length scale $l_P^{2/3}(c/f)^{1/3}$,
characteristic of the noise from spacetime
foam at that frequency, is comparable to the sensitivity level of
the interferometer.  The hope is that, within a certain range of
frequencies, the experimental limits will soon be comparable to
the theoretical predictions for the noise from quantum foam.

The detection of spacetime foam with
interferometry techniques is also helped by the fact that the
correlation length of the noise from spacetime fluctuations is
extremely short, as the characteristic scale is the Planck length.
Thus, this faint noise can be easily distinguished from the other
sources of noise because of this lack of correlation.  In this
regard, it will be very useful for the detection of spacetime foam
to have two nearby interferometers.  

To proceed with the analysis, one first decomposes the 
displacement noise in terms of the associated
displacement amplitude spectral density\cite{radeka}
$S(f)$ of frequency $f$.  For 
the displacement noise due to quantum foam, it is given by
$S(f) \sim c^{1/3} l_P^{2/3} f^{-5/6}$, inversely proportional to
(the $5/6$th power of) frequency.  So one can optimize the 
performance of an interferometer at low frequencies.  As lower 
frequency detection is possible only in space, interferometers like
the proposed Laser Interferometer Space Antenna\cite{Danzmann} 
may enjoy a certain advantage.

To be specific, let us now 
compare the predicted spectal density from quantum foam noise
with the noise level projected for the Laser Interferometer 
Gravitational-Wave Observatory. The ``advanced phase'' of 
LIGO\cite{abram2} is
expected to achieve a displacement noise level of less than $10^{-20}
{\rm mHz}^{-1/2}$ near 100 Hz; 
one can show that this would translate into a
probe of $l_P$ down to
$10^{-31}$ cm, a mere hundred times the physical Planck length.  But can
we then conclude that LIGO will be within striking distance of detecting
quantum foam?
Alas, the above optimistic estimate is based on the assumption that
spacetime foam affects the paths of all the
photons in the laser beam coherently.  
But, in reality, this can hardly be the case.
Since the total effect on
the interferometer is based on averaging over all photons in the wave
front, the incoherent contributions from the different photons
are expected to cut down the sensitivity of the interferometer
by some fractional power of the number of photons in the beam
--- and there are many photons in the beams used by LIGO.
Thus, even with the incredible sensitivity of modern 
gravitational-wave interferometers like LIGO, the fluctuations of
spacetime are too small to be detected --- unless one knows how to
build a small beam interferometer of slightly improved power and phase
sensitivity than what is projected for the advanced phase of 
LIGO!\footnote{This conclusion is based on the author's discussion with
G. Amelino-Camelia and R. Weiss.}

For completeness, we should mention that the use of atom
interferometers\cite{stfoam,Per}
and optical interferometers\cite{group} to look for effects of
spacetime fluctuations has also been suggested.

Last but not least, spacetime foam physics has been applied
to explain some baffling ultra-high energy cosmic
ray events\cite{CRexpt} reported by the Akeno Giant Air Shower Array
observatory in Japan.  But there are
uncertainties on both the observational and theoretical sides.  We
relegate a short discussion on the UHECR events to the Appendix.

\section{Summary and Conclusion}
We summarize by collecting some of the salient points:

\begin{itemize}

\item
On large scales spacetime appears smooth, but on a sufficiently
small
scale it is bubbly and foamy (just as the ocean appears smooth at high
altitudes but shows its roughness at close distances from its surface).

\item
Spacetime is foamy because it undergoes quantum
fluctuations
which give rise to uncertainties in spacetime
measurements; spacetime fluctuations scale as the cube root of 
distances or time durations.

\item
Quantum foam physics is closely related to black hole physics and 
computation.
The ``strange'' holographic principle, 
which limits how densely information
can be packed in space, is a manifestation of quantum foam.

\item
Because the Planck length/time is so small, 
the uncertainties in spacetime measurements, though much greater than
the Planck scale, are still very small.  

\item
It may be difficult to detect the
tiny effects of quantum foam, but it is 
by no means impossible.  

\end{itemize}

Recall that, by
analyzing a simple gedanken experiment for spacetime
measurements, we arrive at the conclusion that
spacetime fluctuations scale as the cube root of
distances or time durations.  This cube root dependence is
mysterious, but has been shown to be consistent with the
holographic principle and with semi-classical black hole physics
in general.  Thus, to this author, this result for spacetime
fluctuations is as beautiful as it is strange (and hopefully also
true)!  Perhaps Sir Francis Bacon was indeed right:
There is no excellent beauty that hath not some strangeness in the
proportion.

But strange beauty is no guarantee for experimental vindication.  What 
is needed is direct detection of quantum foam.  Its detection
will give us a glimpse of the fabric of spacetime and will help
guide physicists to the correct theory of quantum gravity.
The importance of direct experimental evidence cannot be 
over-emphasized.

Now the ball is in the experimentalists' court.


\section*{Acknowledgments}
This work was supported in part by the US Department of Energy and the
Bahnson Fund of the University of North Carolina.  Help from L.~L. Ng
in the preparation of this manuscript is gratefully acknowledged.
I also thank my collaborators, especially H. van Dam and G.
Amelino-Camelia, 
and R. Weiss for useful discussions.  Thanks are due to B. Schumacher
and M. Taqqu for conversations which led to the inclusion of the 
subsection on the various quantum gravity models.

\appendix

\section{Ultra-high Energy Cosmic Ray Events}

\renewcommand{\theequation}{A.\arabic{equation}}

The universe appears to be more transparent to the ultra-high energy
cosmic rays (UHECRs)\cite{CRexpt} 
than expected.\footnote{For the case of (the not-so-well-established) 
TeV-$\gamma$ events, see Ref. 1.}  Theoretically one expects the
UHECRs to interact with the Cosmic Microwave Background
Radiation and produce pions.
These interactions above the threshold energy
should make observations of UHECRs
with $E > 5 {\cdot} 10^{19}$eV (the GZK limit)\cite{GZK} 
unlikely.  Still UHECRs above the GZK limit 
have been observed.
In this appendix, we attempt to explain the
UHECR paradox by arguing\cite{NLOvD}
that energy-momentum uncertainties
due to quantum gravity (significant only for high energy particles
like the UHECRs),
too small to be detected in low-energy
regime, can
affect particle kinematics so as to raise or even eliminate the
energy thresholds, thereby explaining the threshold 
anomaly.\footnote{Unfortunately, we have nothing useful
to say about the origins of these energetic particles per se.}
(For similar or related approaches, see Ref. 29.)

Relevant to the discussion of the UHECR events
is the scattering process in which an energetic
particle
of energy $E_1$ and momentum $\mathbf{p}_1$ collides head-on with a soft
photon of
energy $\omega$ in the production of two energetic particles with
energy
$E_2$, $E_3$ and momentum $\mathbf{p}_2$, $\mathbf{p}_3$.
After taking into account energy-momentum uncertainties,
energy-momentum conservation demands
\begin{equation}
E_1 + \delta E_1 + \omega = E_2 + \delta E_2 +E_3 + \delta E_3,
\label{appen1}
\end{equation}
and
\begin{equation}
p_1 + \delta p_1 - \omega = p_2 + \delta p_2 + p_3 + \delta p_3,
\label{appen2}
\end{equation}
where $\delta E_i$ and $\delta p_i$
($i = 1, 2, 3$) are given by Eqs. (\ref{de}) and (\ref{dp}),
\begin{equation}
\delta E_i = \gamma_i E_i \left(\frac{E_i}{E_P}\right)^{2/3},\,\,\,\,
\delta p_i = \beta_i p_i \left(\frac{p_i}{m_P c}\right)^{2/3},
\label{appen3}
\end{equation}
and we have omitted $\delta \omega$, the contribution from the uncertainty
of $\omega$, because $\omega$ is small.\footnote{We should mention that
we have not found the proper (possibly nonlinear) transformations of
the energy-momentum uncertainties between different reference frames.
Therefore we apply the results only in the frame in which we do the
observations.}

Combining Eq.~(\ref{appen3}) with the modified dispersion
relations\footnote
{The suggestion that the dispersion relation may be modified by quantum
gravity first appeared in Ref. 30.}
Eq. (\ref{moddisp}) for the incoming energetic particle ($i=1$)and the
two outgoing particles ($i=2,3$), and putting $c=1$,
\begin{equation}
E_i^2 - p_i^2 - \epsilon_i p_i^2 \left({p_i \over E_P}\right)^{2/3}
= m_i^2,
\label{appen4}
\end{equation}
we obtain the threshold energy equation
\begin{equation}
E_{th} = p_0 + \tilde{\eta} {1\over 4\omega}
   {E_{th}^{8/3} \over E_P^{2/3}},
\label{appen5}
\end{equation}
where
\begin{equation}
p_0 \equiv \frac{(m_2 + m_3)^2 - m_1^2}{4 \omega}
\label{appen6}
\end{equation}
is the (ordinary) threshold energy if there were no energy-momentum
uncertainties, and
\begin{equation}
\tilde{\eta} \equiv \eta_1 - \frac{\eta_2 m_2^{5/3} +
        \eta_3 m_3^{5/3}}{(m_2 + m_3)^{5/3}},
\label{appen7}
\end{equation}
with
\begin{equation}
\eta_i \equiv 2\beta_i - 2\gamma_i - \epsilon_i.
\label{appen8}
\end{equation}
Note that, in Eq.~(\ref{appen5}), the quantum gravity correction term is
enhanced by the fact that $\omega$ is so small\cite{ACP}
(compared to $p_0$).

Given that all the $\beta_i$'s, the $\gamma_i$'s and the $\epsilon_i$'s are
of order 1 and can be $\pm$, $\tilde{\eta}$ can be $\pm$ (taking on some
unknown Gaussian distribution about zero),
but it cannot be much bigger than 1 in magnitude.
For positive $\tilde{\eta}$, $E_{th}$ is greater than $p_0$.
The threshold energy increases with $\tilde{\eta}$ to ${3\over 2} p_0$ at
$\tilde{\eta} = \tilde{\eta}_{max}$, beyond which
there is no (real) physical solution to Eq.~(\ref{appen5}) (i.e.,
$E_{th}$ becomes complex) and we interpret this as {\it evading} the
threshold cut.\cite{NLOvD}
The cutoff $\tilde{\eta}_{max}$ is actually very small: $\tilde{\eta}_{max}
\sim 10^{-17}$. 
Thus, energy-momentum uncertainties due to
quantum gravity, too small to be detected in low-energy regime, can (in
principle) affect particle kinematics so as to raise or even eliminate
energy thresholds.  Can this be the solution to the UHECR 
threshold anomaly puzzle?
On the other hand, for negative $\tilde{\eta}$, the threshold
energy is less than $p_0$, i.e., a negative $\tilde{\eta}$ {\it lowers}
the threshold energy.\cite{ACNvD,Aloisio2,Huntsville}
Can this be the
explanation of the opening
up of the ``precocious'' threshold in the ``knee'' region?
Curiously, the interpolation between the ``knee'' region and the GKZ limit
may even explain the ``ankle'' region.\cite{br}


It is far too early to call this a success.  
In fact there are some
problems confronting this particular proposal to solve the 
astrophysical puzzle.  The most serious problem is the question
of matter (in)stability\cite{stability} 
because quantum fluctuations in dispersion relations
Eq.~(\ref{appen4}) can {\it lower} as well as raise the 
reaction thresholds.
This problem may force us to entertain 
one or a combination of the following possibilities: (i) The
fluctuations of the energy-momentum of a particle are not completely
uncorrelated (e.g, the fluctuating coefficients $\beta$, $\gamma$, and
$\epsilon$ in Eqs. (\ref{dp}), (\ref{de}), and (\ref{moddisp})
may be related
such that $\eta_i \approx 0$ in Eq.~(\ref{appen8})); (ii) The time scale at
which quantum fluctuations of energy-momentum occur is relatively
short
\footnote{Unfortunately, these two scenarios also preclude the
possibility that energy-momentum uncertainties are the origin of the
threshold anomaly discussed above.  On the positive side, the threshold
anomaly suggested by the present AGASA data may turn out to be false.  
Data from the Auger Project are expected to settle the issue.} 
(compared to the relevant interaction 
or decay times);
(iii) Both ``systematic'' and ``non-systematic'' effects of quantum gravity
are present,\cite{ACNvD}
but the ``systematic'' effects are large enough to overwhelm the
``non-systematic'' effects.

\end{document}